\documentclass[slac_one]{revtex4}
\usepackage{graphicx}
\usepackage{fancyhdr}
\def\Journal#1#2#3#4{{#1} {\bf #2}, #3 (#4)}

\def\NIMA{{\em Nucl. Instrum. Methods} A}

\def\PRL{\em Phys. Rev. Lett.}
\def\PRC{{\em Phys. Rev.} C}

\def\be{\begin{equation}}
\def\ee{\end{equation}}

\begin{document}

\title{THE NEMO-3 EXPERIMENT AND THE SUPERNEMO PROJECT} 

%

\author{L.~Simard, on behalf of the NEMO-3 and the SuperNEMO collaboration}
\affiliation{LAL, Orsay, Universit\'e Paris-Sud 11, France}

\begin{abstract}
The NEMO experiment is investigating the neutrinoless double beta
decay. The NEMO-3 detector is taking data in the
Frejus Underground Laboratory.

 The goal of the SuperNEMO detector is to reach a sensitivity on the
order of $\mathrm{10^{26}}$ year on the half-life of the $\beta \beta
0 \nu$ process. The chosen isotopes for the future detector are $\mathrm{^{82}Se}$ and
$\mathrm{^{150}Nd}$, because of the reduced background. The collaboration has started a 3-year R$\&$D developpement on all
components : tracking detector, calorimeter, source enrichment and
purification, radiopurity measurements.
\end{abstract}

\maketitle

\thispagestyle{fancy}


\section{THE NEMO~3 DETECTOR}\label{subsec:detector}
The NEMO~3 has been taking data since 2003 in the Modane 
underground laboratory located in the Frejus tunnel at the depth of 4800 m w.e. 
Its method of $\beta\beta$-decay study is based on the 
detection of the electron tracks in a tracking device and the 
energy measurement in a calorimeter. 

The detector~\cite{nemo3} has a cylindrical shape.
Thin source foils ($\sim 50~mg/cm^2$) are located
in the middle of the tracking volume surrounded by the calorimeter.
Almost 10kg of enriched $\beta\beta$ isotopes (listed in Table~\ref{tab:t12})
were used to produce the source foils.
The tracking chamber contains 6180 open drift cells operating in the Geiger mode.
It provides a vertex resolution of about 1 cm.
The calorimeter consists of 1940 plastic scintillator blocks with photomultiplier readout.
The energy resolution is 14-17\%/$\sqrt{E}$ FWHM. The time resolution of 250 ps 
allows excellect suppression of the crossing electron background.
A 25~G magnetic field is used for charge identification.

The detector is capable of identifying e$^-$, e$^+$, $\gamma$ and $\alpha$ particles
and allows good discrimination between signal and background events.
\section{NEMO~3 RESULTS}

\subsection{MEASUREMENT OF $2 \nu \beta \beta$ HALF-LIVES}

Measurements of the $2\nu\beta\beta$ decay half-lives were performed for 7 isotopes
available in NEMO~3 (see Table~\ref{tab:t12}). 
New preliminary results based on higher statistics than previously
are presented here 
for two of these isotopes: $^{48}Ca$ and $^{96}Zr$.
\begin{table}[hbt]
\caption{NEMO~3 results of half-life measurement.\label{tab:t12}}
\vspace{0.4cm}
\begin{center}
\begin{tabular}{ |c|c|c|c|c| }
\hline
Isotope& Mass (g) & Q$_{\beta\beta}$ (keV) & Signal/Background & T$_{1/2}$ [$10^{19}$ years]\\
\hline
$^{100}Mo$ &6914& 3034 & 40   & 0.711 $\pm$ 0.002 (stat) $\pm$ 0.054 (syst)~\cite{prl}\\
$^{82}Se$  &932 & 2995 & 4    & 9.6 $\pm$ 0.3 (stat) $\pm$ 1.0 (syst)~\cite{prl}\\
$^{116}Cd$ & 405& 2805 & 7.5  & 2.8 $\pm$ 0.1 (stat) $\pm$ 0.3 (syst)\\
$^{150}Nd$ &37.0& 3367 & 2.8  & $0.920 ^{+0.025}_{-0.022} $(stat) $\pm$ 0.062 (syst)\\
$^{96}Zr$  &9.4 & 3350 & 1.   & 2.3 $\pm$ 0.2 (stat) $\pm$ 0.3 (syst)\\
$^{48}Ca$  &7.0 & 4772 & 6.8  & $4.4 ^{+0.5}_{-0.4} $(stat) $\pm$ 0.4 (syst)\\
$^{130}Te$ &454 & 2529 & 0.25 & 76 $\pm$ 15 (stat) $\pm$ 8 (syst)\\
 \hline
\end{tabular}
\end{center}
\end{table}

\begin{figure}[htb]
\begin{center}
\includegraphics[width=0.24\textwidth,height=3.2cm]{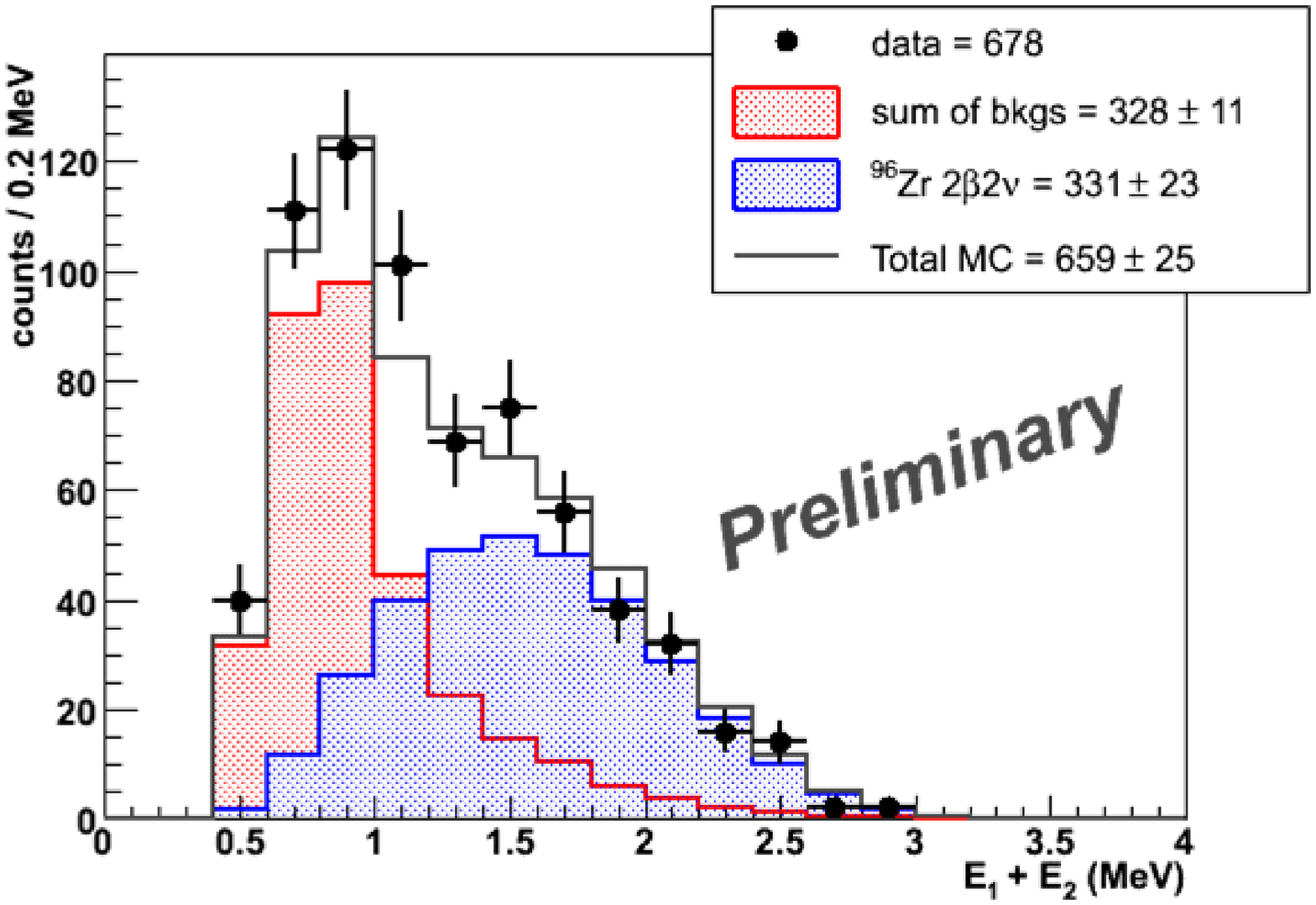}
\includegraphics[width=0.24\textwidth,height=3.2cm]{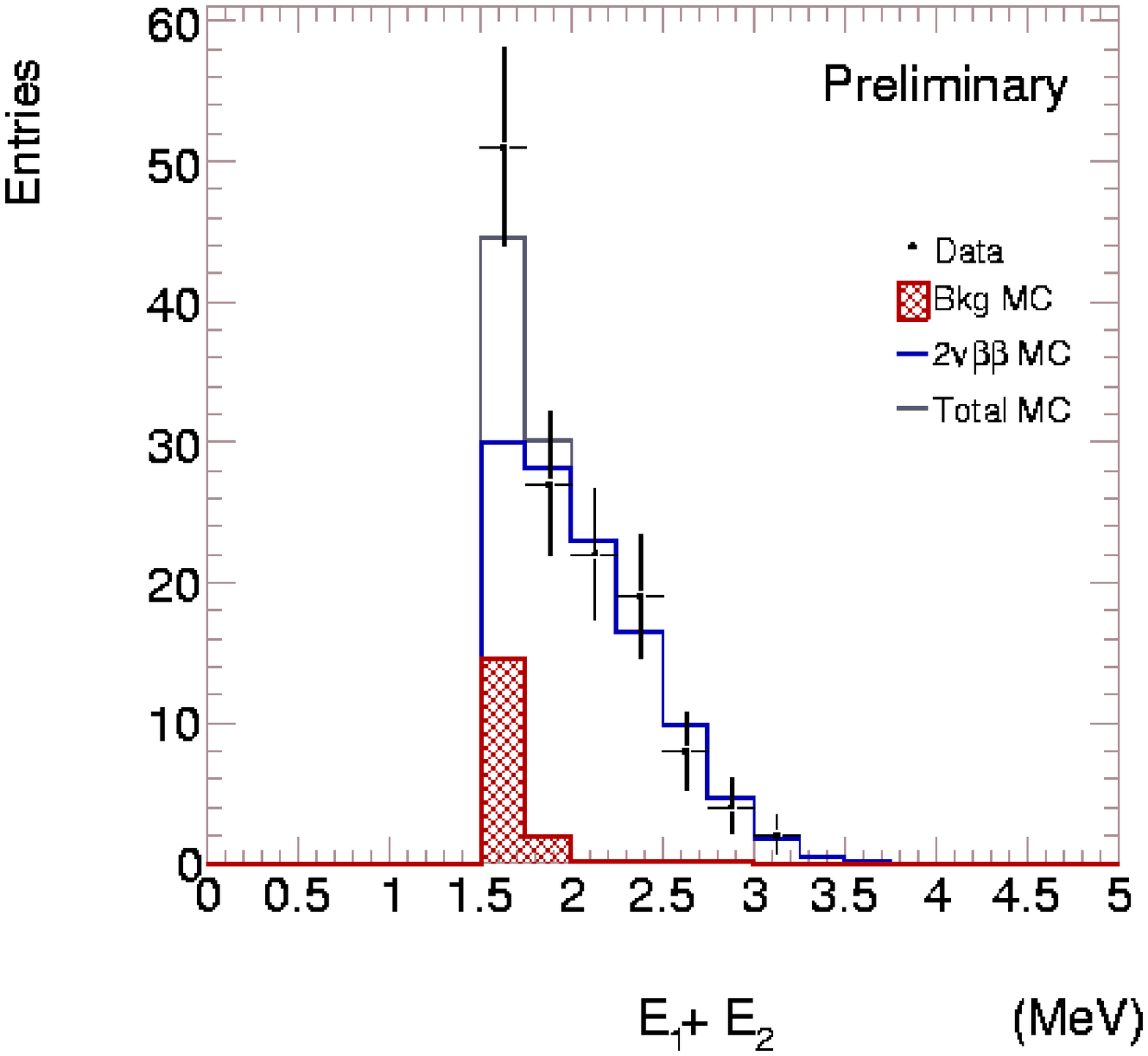}
\end{center}
\caption[Tl]{The energy sum and for two-electron events from $^{96}$Zr
and $^{48}$Ca.}
\label{fig:bb_zr96_ca48}
\end{figure}

The measurement of the $^{96}Zr$ half-life was performed using the data collected within 
925 days. 
A total of 678 events were selected,  
with the expected 328 background events. The largest background
contribution is due to the internal $^{40}$K contamination of the sample. 
The distribution of the energy sum of two electrons and their angular distribution are shown 
in Fig.~\ref{fig:bb_zr96_ca48}, demonstrating good agreement of the data with the Monte Carlo simulation.
The $2\nu\beta\beta$ 
efficiency is estimated to be 7.6\%. The measured half-life is
$T_{1/2}^{2\nu}(^{96}Zr) = [2.3 \pm 0.2 (stat) \pm 0.3 (syst)]\cdot10^{19}y$.

The $^{48}$Ca sample used in NEMO~3 is known to be contaminated by 
$^{90}$Sr(T$_{1/2}$=28.79~y, $Q_{\beta}$=0.546~MeV). Its daughter 
$^{90}$Y (T$_{1/2}$=3.19~h, $Q_{\beta}$=2.282~MeV) is the major background source
in this case. An activity of $1699 \pm 3$ mBq/kg was 
measured for $^{90}$Y  using single-electron events.
Both $^{90}$Sr and $^{90}$Y are essentially pure $\beta^-$ emitters  
and imitate $\beta\beta$ events through M\"oller scattering. 
To supress this background contribution, events with  the energy sum  greater than 1.5 MeV 
and $cos(\Theta_{ee})<0$ are selected. 
Finally, with 943 days of data taking, there are a total of 133 two-electron events, 
with an evaluated residual background contribution of 17 events. 
Their two-electron energy sum distribution are presented in Fig.~\ref{fig:bb_zr96_ca48}.
The $2\nu\beta\beta$ efficiency is 3.3\%, and the measured half-life is 
$T_{1/2}^{2\nu}(^{48}Ca)=[4.4 ^{+0.5}_{-0.4} (stat) \pm 0.4 (syst)]\cdot10^{19}y$.

\subsection{SEARCH FOR $0\nu\beta\beta$ DECAY}

In the case of the mass mechanism, the $0\nu\beta\beta$-decay signal is expected to be a peak 
in the energy sum distribution at the position of the transition energy $Q_{\beta\beta}$.
Since no excess is observed at the tail of the distribution  
for $^{96}$Zr, see Fig.~\ref{fig:bb_zr96_ca48} (left), nor for $^{48}$Ca, Fig.~\ref{fig:bb_zr96_ca48} (right),
limits are set on the neutrinoless double beta decay T$_{1/2}^{0\nu}$ using the CLs method~\cite{cls}.
A lower half-life limit is translated into an 
upper limit on the effective Majorana neutrino mass $\langle m_{\nu}\rangle$.
The following results are obtained using the NME values from papers~\cite{Kort}$^,$ \cite{Simkovic} for $^{96}$Zr
and  from paper~\cite{Caurier}  for $^{48}$Ca
\begin{tabbing}
T$_{1/2}^{0\nu}(^{96}Zr) > 8.6\cdot 10^{21} y$ (90\% C.L.) \hspace{1cm} \= $\langle m_{\nu}\rangle < $ 7.4 -- 20.1 eV \\
T$_{1/2}^{0\nu}(^{48}Ca) > 1.3\cdot 10^{22} y$ (90\% C.L.) \hspace{1cm} \= $\langle m_{\nu}\rangle < $ 29.7 eV.
\end{tabbing} 

\begin{figure}[htb]
\begin{center}
\includegraphics[width=0.24\textwidth,height=3.2cm]{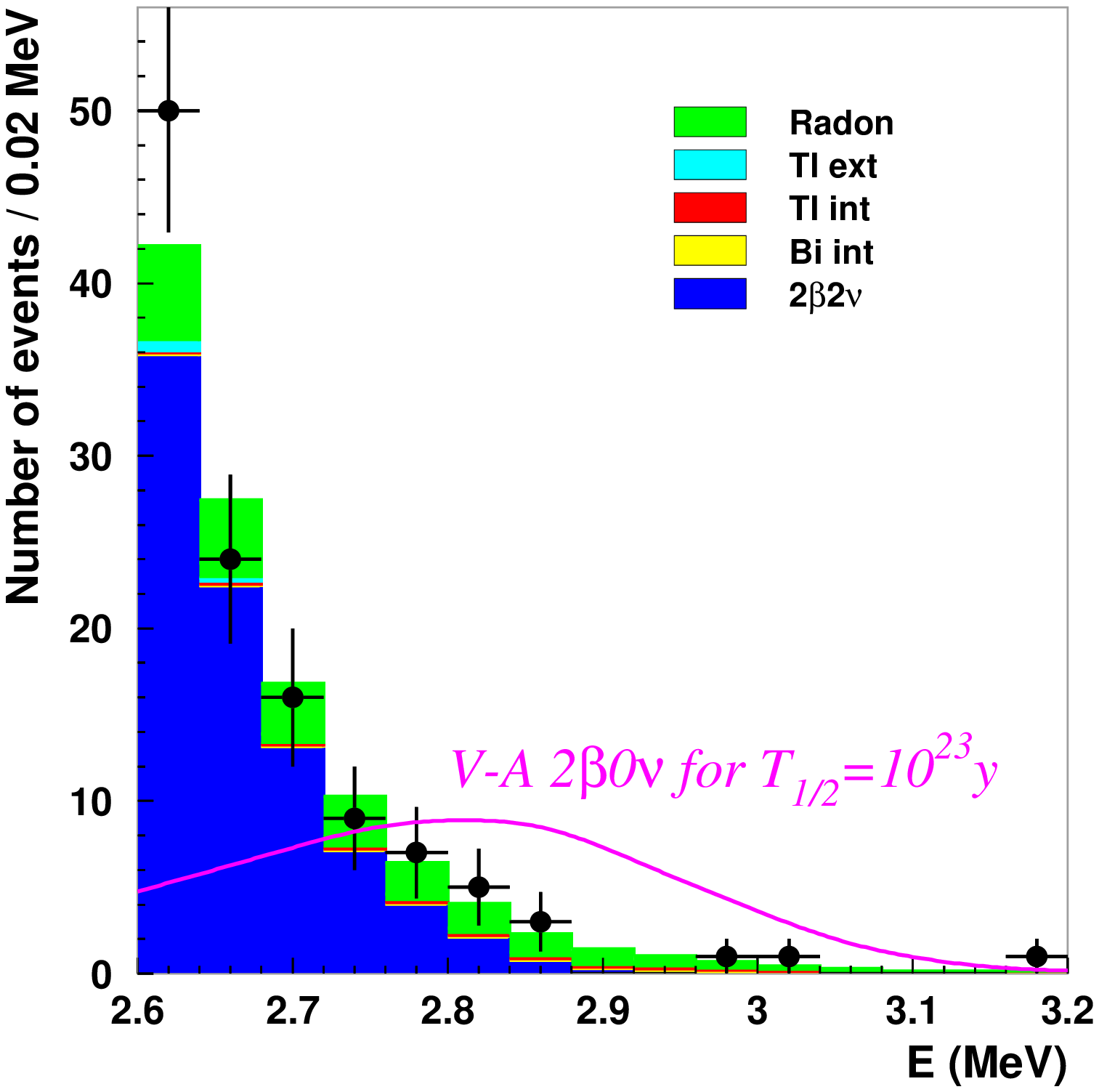}
\includegraphics[width=0.24\textwidth,height=3.2cm]{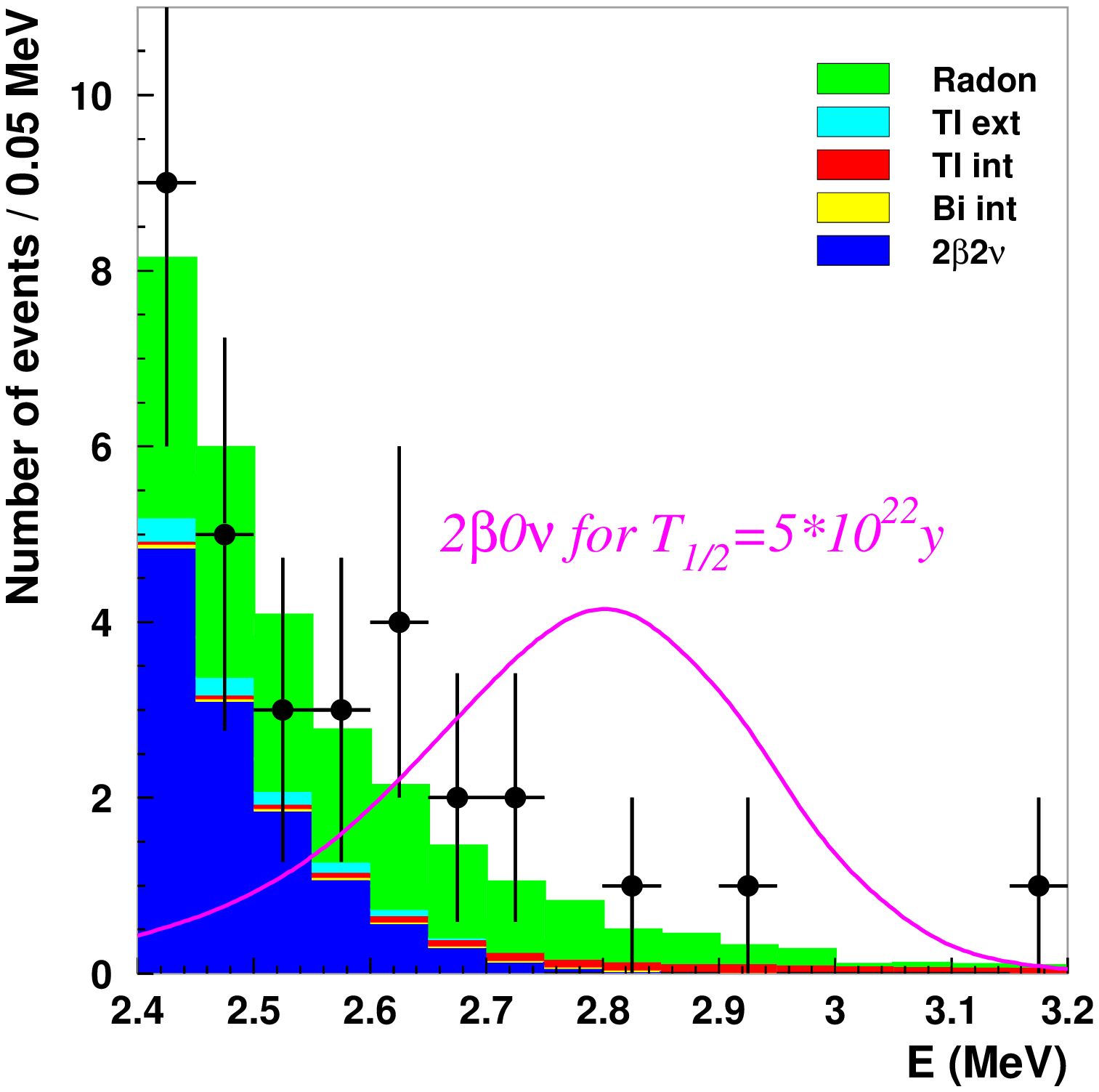}
\end{center}
\caption[Tl]{Distribution of the energy sum of two electrons 
for $^{100}$Mo (left) and  $^{82}$Se (right).}
\label{fig:0nu}
\end{figure}

The $0\nu\beta\beta$-decay search in NEMO~3 is most promising with $^{100}$Mo and $^{82}$Se 
because of the larger available isotope mass and high enough $Q_{\beta\beta} \sim$ 3 MeV.
The two-electron energy sum spectra obtained from the analysis of the data taken within 693 days
are shown in Fig.~\ref{fig:0nu}. For $^{100}$Mo there are 14 events observed in the $0\nu$
search window [2.78-3.20] MeV in good agreement with the expected background of 13.4 events.  
For $^{82}$Se there are 7 events found in the energy sum interval [2.62-3.20] MeV, compared to the expected 
background of 6.4 events. The limits on the T$_{1/2}^{0\nu}$ and the corresponding $\langle m_{\nu}\rangle$
limits calculated using the recent NME values~\cite{Kort}$^,$ \cite{Simkovic} are
\newline
T$_{1/2}^{0\nu}(^{100}Mo) > 5.8\cdot 10^{23} y$ (90\% C.L.) \hspace{0.9cm} $\langle m_{\nu}\rangle <$  0.61--1.26 eV \\
 T$_{1/2}^{0\nu}(^{82}Se) > 2.1\cdot 10^{23} y$ (90\% C.L.) \hspace{1.2cm} $\langle m_{\nu}\rangle <$   1.4 -- 2.2 eV.
\section{THE PRINCIPLE AND CHARACTERISTICS OF THE SUPERNEMO DETECTOR}

The SuperNEMO collaboration has started in February 2006 a 3-year R$\&$D
phase; the goals of this R$\&$D are summarized in table
\ref{tab:1}. This R$\&$D phase has been approved in France, UK and
Spain. Similar proposals are under consideration in Russia, Czech
Republic and Japan. The method is to have R$\&$D tasks on critical
components (obtain a 4 $\%$ FWHM for the calorimeter energy resolution
for 3 MeV electrons, optimize the tracking detector, develop a wiring
automation, produce ultrapure sources and control their purity,
simulate the sensitivity of the detector). At the end of the R$\&$D
phase, a Technical Design Report will be written and the experimental
site (Modane(Frejus), Canfranc, Gran Sasso or Boulby) will be selected.

\begin{table}[hbtp]
\begin{tabular}{rrr}
\hline
  Experiment
  & NEMO-3
  & SuperNEMO \\
\hline
Choice of isotope & $\mathrm{^{100}Mo}$ & $\mathrm{^{150}Nd}$ or $\mathrm{^{82}Se}$ \\

Isotop mass & 7 kg & 100-200 kg \\

Internal contaminations & $\mathrm{^{208}Tl < 20 \mu Bq/kg}$ &
$\mathrm{^{208}Tl < 2 \mu Bq/kg}$ \\

$\mathrm{^{208}Tl}$ and $\mathrm{^{214}Bi}$ in the foil  & $\mathrm{^{214}Bi < 300 \mu Bq/kg}$ & $\mathrm{^{208}Tl < 10 \mu Bq/kg}$ \\

Energy resolution FWHM (calorimeter) & 8$\%$ at 3 MeV   & 4$\%$ at 3 MeV \\

Sensitivity & $\mathrm{T_{1/2}(\beta \beta 0 \nu) > 2 . 10^{24} y}$   & $\mathrm{T_{1/2}(\beta \beta 0 \nu) > 10^{26} y}$ \\

& $\mathrm{<m_\nu> < 0.3-0.9 eV}$ & $\mathrm{<m_\nu> < 40-110 meV}$ \\

\hline

\end{tabular}
\caption{Characteristics of the running experiment NEMO-3 and of the
project of future detector SuperNEMO.}
\label{tab:1}
\end{table}

\begin{figure}[htb]
  \includegraphics[height=3.2cm]{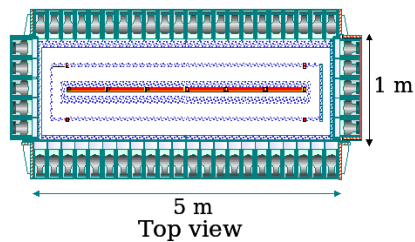}
  \caption{Top view of a module in a possible design of the SuperNEMO
detector. The source shown in red is surrounded by Geiger cells shown
in dark blue and by plastic scintillators blocks coupled to low
radioactivity PMTs.}
  \label{module}
\end{figure}

A possible design for the SuperNEMO detector \cite{FabriceJapon} could be planar and
modular : the 100 kg of enriched isotopes could be placed in 20
modules each containing 5 kg of isotopes. Each source could have a
thickness of 40 $\mathrm{mg/cm^2}$ and a surface of 4 x 3
$\mathrm{m^2}$. For each module, the tracking device could be a drift
chamber made of around 3000 cells, operating in Geiger mode. For each
module, the calorimeter could either be made of around 1000
scintillators blocks coupled to low-radioactivity PMTs, or of
scintillators bars, coupled to around 100 PMTs (see figure \ref{module}).

\section{THE R$\&$D TASKS}

\subsection{R$\&$D ON MEASUREMENT OF MATERIAL RADIOPURITIES}

The best sensitivity for the high-purity Germanium detectors used for
NEMO-3 is 60~$\mu$Bq/kg for $\mathrm{^{208}Tl}$ and 200 $\mu$Bq/kg for
$\mathrm{^{214}Bi}$. The goal of the R$\&$D phase is to improve the
sensitivity by developping 800 $\mathrm{cm^3}$ high purity Germaniun
(with Canberra-Eurisys) with shields improvment and a new ultra-pure
cryostat. A new planar Germanium detector with a resolution of 0.5 keV
at 40 keV is also in development.

The BiPo detector \cite{Mathieu} has been developped to measure the radiopurity in
$\mathrm{^{208}Tl}$ and in $\mathrm{^{214}Bi}$ of the source foils
before their installation in the SuperNEMO detector. The goal is to
measure 5 kg of foils in 1 month with a sensitivity of 2 $\mu$Bq/kg in
$\mathrm{^{208}Tl}$ and of 10~$\mu$Bq/kg in $\mathrm{^{214}Bi}$. The
principle is to tag the electron emitted by the beta desintegration of
$\mathrm{^{212}Bi}$ or of $\mathrm{^{214}Bi}$, then to tag the alpha
emitted by the desintegration of $\mathrm{^{212}Po}$ or of
$\mathrm{^{214}Po}$ (with a decay half-time of 300 ns or of 164
$\mu$s). The thin source can be put in a sandwich of scintillators.
For the measurement of the source contained in one module of
the SuperNEMO detector (12 $\mathrm{m^2}$), the background has to be
very low, less than 1 event per month. Already, a prototype using 20cm
x 20 cm x 3 mm plastic scintillators has been developped and installed
in the Frejus Underground Laboratory; with the
measured background, the expected sensitivity extrapolated for the full BiPo
detector is of the order of 5 $\mu$Bq/kg in $\mathrm{^{208}Tl}$.

\subsection{R$\&$D ON CALORIMETER}

The goal of the calorimeter R$\&$D is to reach a FWHM energy
resolution of 4$\%$ for 3 MeV electrons (7$\%$ for 1 MeV electrons) and to optimize the number of
channels and the detector geometry. 

The goal of the R$\&$D for scintillators is to improve the light yield
and the homogeneity. Plastic scintillators are developped in
collaboration with Kharkhov and Dubna, trying to improve the
performances of polystyrene and to develop polyvinylxylene. Already, a
FWHM of (8.2 $\pm$ 0.1) $\%$ for 1 MeV electrons
has been obtained for 10 cm thick plastic scintillator coupled to a $8^{''}$ PMT. Tests with
different wrappings of the scintillators are also proceeded in
Kharkhov. Liquid
scintillators are also studied : their advantages are the high light
yield, the very good uniformity and transparency; the challenge is to
satisfy the mechanical constraints, especially for the entrance
window, which has to be as thin as possible, in order not to degrade
the electron energy. 

The aim of the R$\&$D for PMTs is to improve the quantum efficiency,
the collection efficiency and to develop low radioactivity PMTs : an
agreement between PHOTONIS and IN2P3 has been signed, tests are also
done with Hamamatsu and ETL. Already, PMTs with very high quantum
efficiencies (43$\%$ for 3 inches and 35$\%$ for 8 inches) have been
developped. Slow and fast PMTs are also studied. The goal is also to have a higher compacity by reducing the number of
channels, without reducing too much the light collection. The energy
measurement with scintillator bars with 2 PMTs or with optical fibers is also studied.
%



\end{document}